\documentclass[twocolumn,prb,aps,10pt]{revtex4-1}

\usepackage{amsfonts,amssymb}
\usepackage[sumlimits,intlimits]{amsmath}
\usepackage{graphics}
\usepackage{graphicx}
\usepackage{epsfig}
\usepackage{mathrsfs}
\usepackage{textcomp}
\usepackage{verbatim}
\usepackage{hyperref}    
\usepackage{bm}          

\newcommand{\be}{\begin{equation}}
\newcommand{\ee}{\end{equation}}
\newcommand{\sgn}{\operatorname{sgn}}
\newcommand{\rr}{{\mathbf{r}}}
\newcommand{\RR}{{\mathbf{R}}}
\newcommand{\Gt}{\widetilde{\Gamma}}
\newcommand{\mt}{\widetilde{\mu}}
\newcommand{\rt}{\widetilde{r}_s}
\newcommand{\kb}{\kappa_\text{b}}
\newcommand{\pext}{\phi_\text{ext}}
\newcommand{\pnew}{\phi_\text{new}}

\newcommand{\phit}{\widetilde{\phi}_1}
\newcommand{\smin}{\sigma_\text{min}}

\begin{document}

\title{Theory of the random potential and conductivity at the surface of a topological insulator}

\author{Brian Skinner}
\author{B. I. Shklovskii}
\affiliation{Fine Theoretical Physics Institute, University of Minnesota, Minneapolis, MN 55455, USA}

\date{\today}

\begin{abstract}

We study the disorder potential induced by random Coulomb impurities at the surface of a topological insulator (TI).  We use a simple model in which positive and negative impurities are distributed uniformly throughout the bulk of the TI, and we derive the magnitude of the disorder potential at the TI surface using a self-consistent theory based on the Thomas-Fermi approximation for screening by the Dirac mode.  Simple formulas are presented for the mean squared potential both at the Dirac point and far from it, as well as for the characteristic size of electron/hole puddles at the Dirac point and the total concentration of electrons/holes that they contain.  We also derive an expression for the autocorrelation function for the potential at the surface and show that it has an unusually slow decay, which can be used to verify the bulk origin of disorder.  The implications of our model for the electron conductivity of the surface are also presented.

\end{abstract}
\maketitle

\section{Introduction}
\label{sec:intro}

The three-dimensional (3D) topological insulator (TI)~\cite{Fu2007tii, Moore2007tio, Roy2009tpa, Fu2007tiw, Qi2008tft} has gapless surface states with a Dirac-like spectrum, which host a number of interesting quantum transport phenomena~\cite{Hasan2010c:t, Qi2011tia}.  These surface states are influenced by the presence of a random Coulomb potential, which is believed to determine the mobility of surface electrons~\cite{Xiong2012hsh}. Recently, the random potential at the surface of typical TIs was studied directly by spectroscopic mapping with a scanning tunneling microscope~\cite{Beidenkopf2011sfh}.  It was shown that near the Dirac energy random fluctuations of the potential have a Gaussian-like distribution with a width $\sim 20$ -- $40$ meV.  For a theoretical interpretation of their results, Ref.~\onlinecite{Beidenkopf2011sfh} used a model of random charges with two-dimensional (2D) concentration $n_i$ situated in a plane parallel to the surface at a distance $d$ from it~\cite{Adam2007sct, Fogler2009npg} that is self-consistently screened by the electrons of the surface Dirac mode. Originally, this model was suggested to describe disorder in graphene, where random charges can be assumed to be localized on the surface of a nearby substrate.  It has since been extended to describe the surface of a 3D TI \cite{Li2012tde}.

In this paper we explore a different model of Coulomb disorder in TIs. We assume that the bulk of the TI is a completely compensated semiconductor with equal 3D concentration $N$ of donors and acceptors, which are randomly-distributed throughout the bulk. This model is mathematically simpler than the 2D one~\cite{Adam2007sct, Fogler2009npg} because impurities are characterized by only one parameter, $N$, rather than two parameters, $n_i$ and $d$.  An analysis of this 3D model is presented below, but before turning to it we would like to give arguments for the 3D model that are specific for known TIs.

First, such a model is justified by current methods of preparation of TI crystals. Typically, as-grown TI crystals are heavily doped $n$-type semiconductors with $N \sim 10^{19}$ donors per cm$^{3}$. The Fermi level of such a crystal is high in the conduction band.  In order to bring the Fermi level down to the middle of the gap and increase the bulk resistivity, the TI is compensated by acceptors with concentration close to that of the donors, $N$.  Below we assume that these donors and acceptors are randomly distributed in space.  This is indeed usually the case for samples made by cooling from a melt, where the distribution of impurities in space is a snapshot of the distribution at much higher temperature, when diffusion of impurities practically freezes~\cite{Keldysh1964}. In semiconductors with a narrow enough forbidden gap $E_g$, at this temperature there is a concentration of intrinsic carriers larger than the concentration of impurities.  These intrinsic carriers screen the Coulomb interaction between impurities, so that the impurities remain randomly distributed in space.  As the melt is cooled to the point where intrinsic carriers recombine, the impurities are left in random positions~\cite{Galpern1972epc, Efros1984epo}.  If diffusion freezes at $T \sim 1000$\,K it is reasonable to assume that impurities are randomly positioned in a semiconductor with $E_g \leq 0.3$ eV. 

Second, the model of a completely compensated TI with randomly positioned charges was recently tested by calculation of the activation energy $\Delta$ of its bulk resistivity and comparison to experiment~\cite{Skinner2012wib}.  The standard expectation, which assumes flat valence and conduction bands, was that when the Fermi level is moved to the middle of the gap the bulk of the TI becomes a good insulator with activation energy $\Delta = E_g/2$. In reality, in the Bi$_2$Se$_3$ and Bi$_2$Te$_3$ families of TIs, where $E_g \sim 0.3$\,eV, the activation energy $\Delta$ was found~\cite{Ren2011o$s} to be frustratingly small, with $\Delta \approx 0.15 E_g$. This unexpectedly small activation energy was shown to be explainable within the model of random 3D donor and acceptor charges.  Specifically, it was shown by numerical simulation~\cite{Skinner2012wib} that $\Delta \approx 0.15 E_g$ results from band bending by the potential created by random 3D Coulomb impurities. For these reasons we consider our model of 3D randomly situated donors and acceptors to be an appropriate description of the TI bulk.  

Our primary result for this model is an expression for the amplitude of fluctuations of the electric potential energy, $\Gamma$, at the TI surface as a function of the chemical potential, $\mu$, measured relative to the Dirac point.  In particular, for $\mu = 0$ we show below that 
\be 
\Gamma^2 = \frac{ \sqrt[3]{2} \pi}{\alpha^{4/3}}  \left( \frac{e^2 N^{1/3}}{\kappa} \right)^2,  \hspace{5mm} (\mu = 0).
\label{eq:gamma0}
\ee 
Here, $-e$ is the electron charge, $\kappa$ is the effective dielectric constant, and $\alpha = e^2/\kappa \hbar v$ is the effective fine structure constant, where $\hbar$ is the reduced Planck constant and $v$ is the Dirac velocity.  This expression describes screening of the disorder potential via the formation of electron and hole puddles at the TI surface.  The characteristic size of these puddles is given by
\be 
r_s = \frac{N^{-1/3}}{2^{2/3} \alpha^{4/3}},  \hspace{5mm} (\mu = 0),
\label{eq:rs0units}
\ee
and the corresponding total number of electrons (or holes) per unit area in surface puddles is given by
\be 
n_p = \left( \frac{\alpha}{16} \right)^{2/3} N^{2/3},  \hspace{5mm} (\mu = 0).
\label{eq:ne}
\ee
Eqs.\ (\ref{eq:gamma0}) -- (\ref{eq:ne}) are derived below, along with results corresponding to large $\mu$.  Our result for the mean squared potential, $\Gamma^2$, is plotted in Fig.\ \ref{fig:p2vsmu} as a function of $\mu$.  Below we also derive a simple relation for the autocorrelation function of the potential at the TI surface, which has an unusually slow decay and can be used to verify the bulk origin of disorder.

In addition to describing the disorder potential, we calculate the corresponding electron conductivity $\sigma$ of the surface, and we show that when the average electron concentration $n$ satisfies $n \gg n_p$, the conductivity is given by
\be
\sigma \simeq \frac{e^2}{h} \frac{2 \sqrt{\pi}}{\alpha^2 \ln(1/\alpha)} \frac{n^{3/2}}{N},
\label{eq:sigman}
\ee
where $e^2/h$ is the conductance quantum.  At much smaller electron concentrations, $n \ll n_p$, the conductivity saturates at a value $\smin$, which we estimate as
\be 
\smin \simeq \frac{e^2}{h} \frac{1}{\pi \alpha \ln(1/\alpha)}.
\label{eq:smin}
\ee

The remainder of this paper is organized as follows.  In Sec.\ \ref{sec:theory} we develop our self-consistent theory to describe the screened disorder potential at the TI surface.  In Sec.\ \ref{sec:sim} these analytical results are compared to results from a numerical simulation of the TI surface.  Sec.\ \ref{sec:conductivity} presents the implications of our model for the electron conductivity of the surface.  We conclude in Sec.\ \ref{sec:discussion} by comparing our theory with the recent experiments of Ref.\ \onlinecite{Beidenkopf2011sfh}, and by discussing the major assumptions of our theory and its implications for future experiments.

\section{Self-consistent theory of the surface disorder potential}
\label{sec:theory}

In the limit where the potential varies slowly compared to the characteristic Fermi wavelength of electrons at the surface, the electric potential $\phi(\rr)$ can be described using the Thomas-Fermi (TF) approach:
\be 
\mu = E_f[n(\rr)] - e \phi(\rr).
\label{eq:TF}
\ee 
Here, $E_f(n) = \hbar v \sqrt{4 \pi |n|}\sgn(n) = (e^2/\alpha \kappa)\sqrt{4\pi |n|} \sgn(n)$ is the local Fermi energy and $n(\rr)$ is the 2D electron concentration at the point $\rr$ on the surface.  The TF approximation is justified whenever $\alpha \ll 1$, as we show below.  In TIs such small $\alpha$ can be seen as the result of the large bulk dielectric constant $\kb \gtrsim 30$.  We note here that for describing the properties of the surface state, which exists  at a dielectric discontinuity, one should use for the effective dielectric constant $\kappa$ the arithmetic mean of the internal and external dielectric constants.  If the TI is in vacuum, then $\kappa = (\kb + 1)/2 \simeq \kb/2$.  

When the chemical potential is large enough in magnitude that $\mu^2 \gg e^2 \langle \phi^2 \rangle$, where $\langle ... \rangle$ denotes averaging over the TI surface, the relation $E_f(n)$ can be linearized to read $E_f[n(\rr)] \simeq \mu + \delta n(\rr)/\nu(\mu)$.  Here $\delta n(\rr) = n(\rr) - n_0$ is the difference in the electron concentration relative to the state with zero electric potential, $n_0 = \alpha^2 \kappa^2 \mu^2/(4 \pi e^4)$, and $\nu(\mu) = \alpha^2 \kappa^2 |\mu|/(2 \pi e^4)$ is the density of states at $E_f = \mu$.  From this density of states one can define a screening radius $r_s = \kappa/2\pi e^2 \nu = e^2/\alpha^2 \kappa \mu$ that characterizes the distance over which fluctuations in the Coulomb potential are screened by the surface.  The TF approximation is valid when the Fermi wavelength $\lambda_f \sim n_0^{-1/2} \sim e^2/\alpha \kappa \mu$ is much smaller than $r_s$, which gives the condition $\alpha \ll 1$.

One can understand qualitatively the magnitude of the potential fluctuations, $\Gamma$, using the following simple argument.  For a given point on the TI surface, one can say that only impurities within a distance $R \lesssim r_s$ contribute to the potential; those impurities at a distance $R \gg r_s$ are effectively screened out (one can say that they are screened by their image charges in the ``metallic" TI surface).  Impurities with $R < r_s$, on the other hand, are essentially unscreened.  There are $\sim N r_s^3$ such impurities, and their net charge is of order $Q \sim e \sqrt{N r_s^3}$, with a random sign.  The absolute value of the potential at the surface is then $\sim Q/\kappa r_s$, so that $\Gamma \sim e Q/\kappa r_s \sim (e^2 N^{1/3}/\kappa) (N r_s^3)^{1/6} \sim \sqrt{e^2 N/\kappa \nu} \sim \sqrt{e^4 N/\alpha^2 \kappa^3 |\mu|}$.  Throughout this paper we assume that all donors and acceptors in the bulk are ionized, and we ignore the possible effects of bulk screening.  This assumption is generally justified as long as the bulk chemical potential resides in the band gap, as we discuss in Sec.\ \ref{sec:discussion}.

In order to more accurately derive the value of $\Gamma$, one can start by considering the potential created by a single impurity charge $+e$.  When such an impurity charge is placed a distance $z$ from the TI surface (say, above the origin), it creates a potential $\phi_1(r; z)$ that within the TF approximation is given by \cite{Ando1982epo}
\be 
\phi_1(r; z) = \frac{e}{\kappa} \int_0^\infty \frac{\exp [-q z]}{1 + (q r_s)^{-1}} J_0(q r) \, dq ,
\label{eq:p1}
\ee 
where $J_0(x)$ is the zeroth order Bessel function of the first kind.  At large $z/r_s$, Eq.\ (\ref{eq:p1}) can be expanded to give 
\be 
\phi_1(r; z) \simeq \frac{e}{\kappa} \frac{z r_s}{(r^2 + z^2)^{3/2}}.
\label{eq:p1large}
\ee
A simple physical derivation of Eq.\ (\ref{eq:p1large}) is based on the notion \cite{Loth2009nsb} that for a distant impurity, such that $z \gg r_s$, a surface with screening radius $r_s$ effectively plays the role of a metallic surface positioned below the real surface at a distance $z = -r_s/2$.  Equation (\ref{eq:p1large}) can then be viewed as the sum of the potentials created by the original charge at a distance $z$ above the plane and its opposite image charge at a distance $z + r_s$ below the plane, expanded to lowest order in $r_s/z$.

The total potential at the origin is $\phi(0) = \sum_i q_i \phi_1(r_i; z_i)$, where the index $i$ labels all impurity charges, $q_i$ is the sign of impurity $i$, and $\rr_i$ and $z_i$ are the radial and azimuthal coordinates of its position.  Under the assumption that all impurity positions are uncorrelated and randomly-distributed throughout the bulk of the TI, the average of $\phi^2$ is given by
\be 
\langle \phi^2 \rangle = \int [\phi_1(r'; z')]^2 \, 2 N d^2 \rr' dz'.
\label{eq:gammaint}
\ee 
Here, the quantity $2N d^2\rr' dz'$ describes the probability that the volume element $d^2\rr' dz'$ contains an impurity charge, and the integration is taken over the semi-infinite volume of the bulk of the TI.  The width of the disorder potential at the TI surface, $\Gamma$, is defined by $\Gamma^2 = e^2 \langle \phi^2 \rangle$.  Inserting Eq.\ (\ref{eq:p1}) into Eq.\ (\ref{eq:gammaint}) and taking the integral then gives
\be 
\Gamma^2 = \frac{e^2 N}{\kappa \nu} = \frac{2 \pi e^4 N}{\alpha^2 \kappa^3 |\mu|}, \hspace{5mm} \left( |\mu| \gg \frac{e^2 N^{1/3}}{\kappa \alpha^{2/3}} \right). \label{eq:gammalargemu}
\ee 
Eq.\ (\ref{eq:gammalargemu}) is correct so long as the fluctuations in the Coulomb potential energy are small compared to the chemical potential, or $\Gamma \ll |\mu|$; this gives the condition written in parentheses.

On the other hand, when $|\mu|$ is very small, the fluctuations in the Coulomb potential become large compared to the chemical potential, and one cannot talk about a constant density of states $\nu$ or screening radius $r_s$.  Instead, the Fermi energy has strong spatial variations, and the random potential is screened by the formation of electron and hole puddles at the surface.  Nonetheless, one can define an average density of states $\langle \nu \rangle$ at the surface, which determines, self-consistently, the typical screening radius $r_s$ and the magnitude of the potential fluctuations at the TI surface.

Consider, for example, the case $\mu = 0$, where by symmetry the average value of the potential $\langle \phi \rangle = 0$.  At any given point $\rr$ on the surface, the potential $\phi(\rr)$ is the sum of contributions from many individual impurity charges, provided that the characteristic screening radius $r_s = \kappa/2\pi e^2 \langle \nu \rangle \gg N^{-1/3}$.  This implies that, by the central limit theorem, the value of the potential across the surface is  Gaussian-distributed with some variance $\langle \phi^2 \rangle = \Gamma^2/e^2$ that remains to be calculated.  Within the TF approximation the local density of states at the point $\rr$ is $\nu[-e\phi(\rr)] = e \alpha^2 \kappa^2 |\phi(\rr)|/(2 \pi e^4)$, so that one can calculate the average density of states as
\begin{eqnarray} 
\langle \nu \rangle & = & \int_{-\infty}^{\infty} \nu(-e\phi) \frac{\exp\left[ -e^2\phi^2/2 \Gamma^2 \right]}{\sqrt{2 \pi \Gamma^2/e^2}}  \, d\phi \nonumber \\
& = & \frac{\alpha^2\kappa^2 \Gamma}{\sqrt{2 \pi^3} e^4},  \hspace{5mm} (\mu = 0).
\label{eq:nu0}
\end{eqnarray}
This result for $\langle \nu \rangle$ can be inserted into the first equality of Eq.\ (\ref{eq:gammalargemu}), $\Gamma^2 = e^2 N/\kappa \langle \nu \rangle$, to give a self-consistent relation for the amplitude of potential fluctuations \cite{Stern1974ltl}.  This procedure gives the result first announced in the Introduction, Eq.\ (\ref{eq:gamma0}).
Substituting Eqs.\ (\ref{eq:gamma0}) and (\ref{eq:nu0}) into the expression for the screening radius, $r_s = \kappa/2\pi e^2 \langle \nu \rangle$, gives Eq.\ (\ref{eq:rs0units}).

One can also calculate the total concentration of electrons/holes in surface puddles, $n_p$, implied by this result for $\Gamma^2$.  This is done by first inverting the TF relation, Eq.\ (\ref{eq:TF}), at $\mu = 0$ to give $n(\phi) = (\alpha^2 \kappa^2/4 \pi e^2) \phi^2 \sgn(\phi)$.  Integrating this expression for $n(\phi)$ weighted by the Gaussian probability distribution for $\phi$ gives
\begin{eqnarray} 
n_p & = & \int_0^{\infty} n(\phi) \frac{\exp\left[ -e^2\phi^2/2 \Gamma^2 \right]}{\sqrt{2 \pi \Gamma^2/e^2}} d \phi \nonumber \\
& = & \frac{\alpha^2 \kappa^2 \Gamma^2}{8 \pi e^4},  \hspace{5mm} (\mu = 0). \nonumber
\nonumber 
\end{eqnarray}
Substituting the result of Eq.\ (\ref{eq:gamma0}) for $\Gamma^2$ then gives Eq.\ (\ref{eq:ne}).  One can also combine this result for the residual electron/hole concentration, $n_p$, with the expression for the screening radius, $r_s$, to arrive at an estimate for the number of electrons/holes per puddle: $M_p \sim \pi n_p r_s^2 \sim \pi/16 \alpha^2$.  Apparently at small $\alpha$ puddles typically contain many electrons/holes, $M_p \gg 1$.

Our primarily results, outlined in Eqs.\ (\ref{eq:gamma0}) -- (\ref{eq:ne}), are valid within the TF approximation so long as the typical Fermi wavelength, $\lambda_f \sim e^2/\alpha \kappa \Gamma$, is much smaller than the typical screening radius, $r_s \sim  e^2/\alpha^2 \kappa \Gamma$, which again gives the condition $\alpha \ll 1$.  Equations (\ref{eq:gamma0}) and (\ref{eq:rs0units}) were obtained in Ref.\ \onlinecite{Shklovskii2007smc} without numerical coefficients within the context of graphene on a silicon oxide substrate.

One can notice that our expressions for $\Gamma^2$ can be expressed more compactly by defining the dimensionless units $\Gt = \Gamma/E_0$ and $\mt = \mu/E_0$, where $E_0 = e^2 N^{1/3}/\alpha^{2/3} \kappa$.  In these units Eqs.\ (\ref{eq:gamma0}) and (\ref{eq:gammalargemu}) can be written as
\be 
\Gt^2 = \sqrt[3]{2} \pi \approx 3.96, \hspace{5mm} (\mt = 0 )
\label{eq:gs0}
\ee
and
\be 
\Gt^2 = 2 \pi/|\mt|, \hspace{5mm} (|\mt| \gg 1)
\label{eq:gslarge},
\ee 
respectively, and the constant $\alpha$ does not enter explicitly.  Eqs.\ (\ref{eq:gs0}) and (\ref{eq:gslarge}) are plotted as the red dotted and dashed lines, respectively, in Fig.\ \ref{fig:p2vsmu}.  One can similarly define a dimensionless screening radius $\rt = r_s/r_0$, where $r_0 = N^{-1/3}/\alpha^{4/3}$, so that
\be 
\rt = 2^{-2/3} \approx 0.63, \hspace{5mm} (\mt = 0 )
\label{eq:rs0}
\ee
and
\be 
\rt = 1/|\mt|, \hspace{5mm} (|\mt| \gg 1)
\label{eq:rslarge}.
\ee 

At $\mu = 0$, the screening radius $r_s$ describes the characteristic size of electron or hole puddles at the TI surface.  More generally, $r_s$ plays the role of a length scale over which potential fluctuations at the surface are correlated.  Such correlations can be discussed in a quantitative way by defining the potential auto-correlation function:
\be 
C(r) = \langle \phi(\RR) \phi(\RR + \rr) \rangle_\RR ,
\label{eq:Cdef}
\ee
where $\langle ... \rangle_\RR$ denotes averaging over the spatial coordinate $\RR$, and where by symmetry the correlation function depends on $|\rr| = r$ only.  Before proceeding to present numerical results, we first derive approximate analytical results for $C(r)$, and show that spatial correlations in the potential have an unusually slow decay.

At $r = 0$, Eq.\ (\ref{eq:Cdef}) reproduces the expression for $\langle \phi^2 \rangle$, so that $C(0) = \Gamma^2/e^2$.  At small enough distances that $r \ll r_s$, one can expect that the value of $C(r)$ is determined primarily by unscreened impurities that are within a distance $r_s$ from the surface, as explained above during the derivation of $\Gamma^2$.  On the other hand, at $r \gg r_s$ correlations are produced primarily by impurities that are relatively far from the surface, as can be seen from the following scaling argument.  Consider two surface points separated by a distance $r \gg r_s$.  One can imagine drawing a cube of size $r$ that extends into the bulk of the TI and which contains the two surface points on opposite edges of one of its faces.  Such a cube contains $\sim N r^3$ impurities, and has a net impurity charge with magnitude $q \sim e \sqrt{N r^3}$ and random sign.  These impurity charges are located at a mean distance $\sim r \gg r_s$ above the surface and, therefore, by Eq.\ (\ref{eq:p1large}), contribute a net potential $\sim q r_s /\kappa r^2 \sim (e/\kappa) \sqrt{N r_s^2/r}$ to both surface points.  The square of this potential roughly gives the autocorrelation of the potential, $C(r) \sim e^2 Nr_s^2 /\kappa^2 r$.

A more careful expression for $C(r)$ can be derived by writing
\be 
C(r) = \int \phi_1(\rr'; z') \phi_1(\rr'-\rr; z') 2N d^2\rr' dz',
\label{eq:Cint}
\ee 
similar to Eq.\ (\ref{eq:gammaint}).  Inserting the asymptotic expression of Eq.\ (\ref{eq:p1large}) for $\phi_1$ and evaluating the integral gives
\be 
C(r) \simeq \frac{2 \pi e^2 N r_s^2}{\kappa^2 r} = \frac{\Gamma^2/e^2}{r/r_s}, \hspace{5mm} (r/r_s \gg 1).
\label{eq:C}
\ee
This result is plotted as the dashed line in Fig.\ \ref{fig:C}.  

Eq.\ (\ref{eq:C}) implies an unusually slow decay of potential correlations at the surface, which, as explained above, arises from long-range fluctuations of the potential created by deep bulk impurities.  This behavior can be contrasted with the much faster decay of $C(r)$ that would result from the 2D model of planar impurities
\footnote{
This result can be obtained by replacing the bulk impurity charge density $2N$ in Eq.\ (\ref{eq:Cint}) with $n_i \delta(z-d)$.
}: $C(r) \sim e^2 n_i d r_s^2 / \kappa^2 r^3$.  Thus, by studying $C(r)$ experimentally by scanning tunneling microscopy, one can discriminate between disorder by bulk impurities and disorder by impurities located in a layer close to the surface.

\section{Numeric simulation}
\label{sec:sim}

So far we have presented analytical results for the magnitude of the potential at the surface and its autocorrelation function.  These results are derived within the approximation of linear screening with a self-consistent screening radius $r_s$.  One can question the accuracy of this approach, particularly at small $\mt$, where the density of states varies strongly from one point to another.  Therefore, in order to verify the analytical results presented above, we implemented a simple simulation of a TI surface with adjacent point-like impurities and we solved numerically for the electric potential $\phi$ at arbitrary $\mt$ within the TF approximation.  In these simulations, a square planar surface of dimension $L \times L$ is placed adjacent to a volume of size $L \times L \times L/2$ with $NL^3$ randomly-positioned impurities, each with a random sign.  These impurities create a bare potential $\pext(\rr)$ at the surface, and the self-consistent potential $\phi(\rr)$ satisfies
\be 
\phi(\rr) = \pext(\rr) - \int d^2 \rr' \frac{ e \delta n(\rr')}{\kappa | \rr - \rr' |} \nonumber .
\ee
The TF equation, Eq.\ (\ref{eq:TF}), can be inverted to read
\be 
\delta n(\rr) = \frac{ \alpha^2 \kappa^2}{4 \pi e^3} \left| 2 \mu \phi(\rr) + e \left[ \phi(\rr) \right]^2 \right| \sgn[\phi(\rr)] \nonumber ,
\ee
so that the self-consistent equation for the potential $\phi(\rr)$ can be written 
\be 
\phi(\rr) = \pext(\rr) - \frac{ \alpha^2 \kappa}{4 \pi e^2} \int d^2 \rr' \frac{ \left| 2 \mu \phi(\rr') + e \left[ \phi(\rr') \right]^2 \right|}{| \rr - \rr' |} \sgn[\phi(\rr')] .
\label{eq:selfconsistent}
\ee

\begin{figure}[htb!]
\centering
\includegraphics[width=0.5 \textwidth]{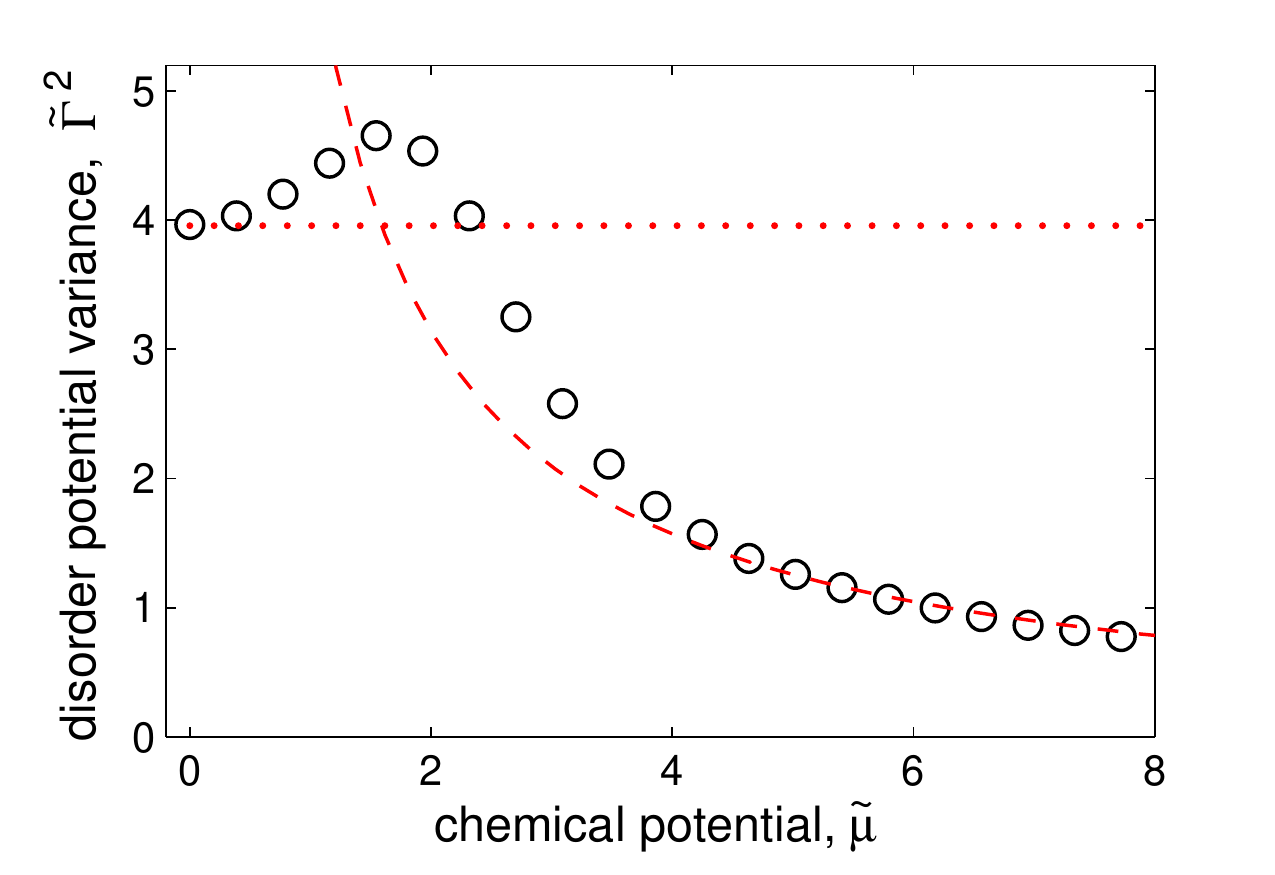}
\caption{(Color online)  Variance in the disorder potential at the TI surface as a function of the chemical potential relative to the Dirac point.  The dotted and dashed lines correspond to Eqs.\ (\ref{eq:gs0}) and (\ref{eq:gslarge}), respectively.  Open circles show the result of a numerical solution of Eq.\ (\ref{eq:selfconsistent}); error bars are smaller than the symbol size.}
\label{fig:p2vsmu}
\end{figure}

\begin{figure}[htb!]
\centering
\includegraphics[width=0.5 \textwidth]{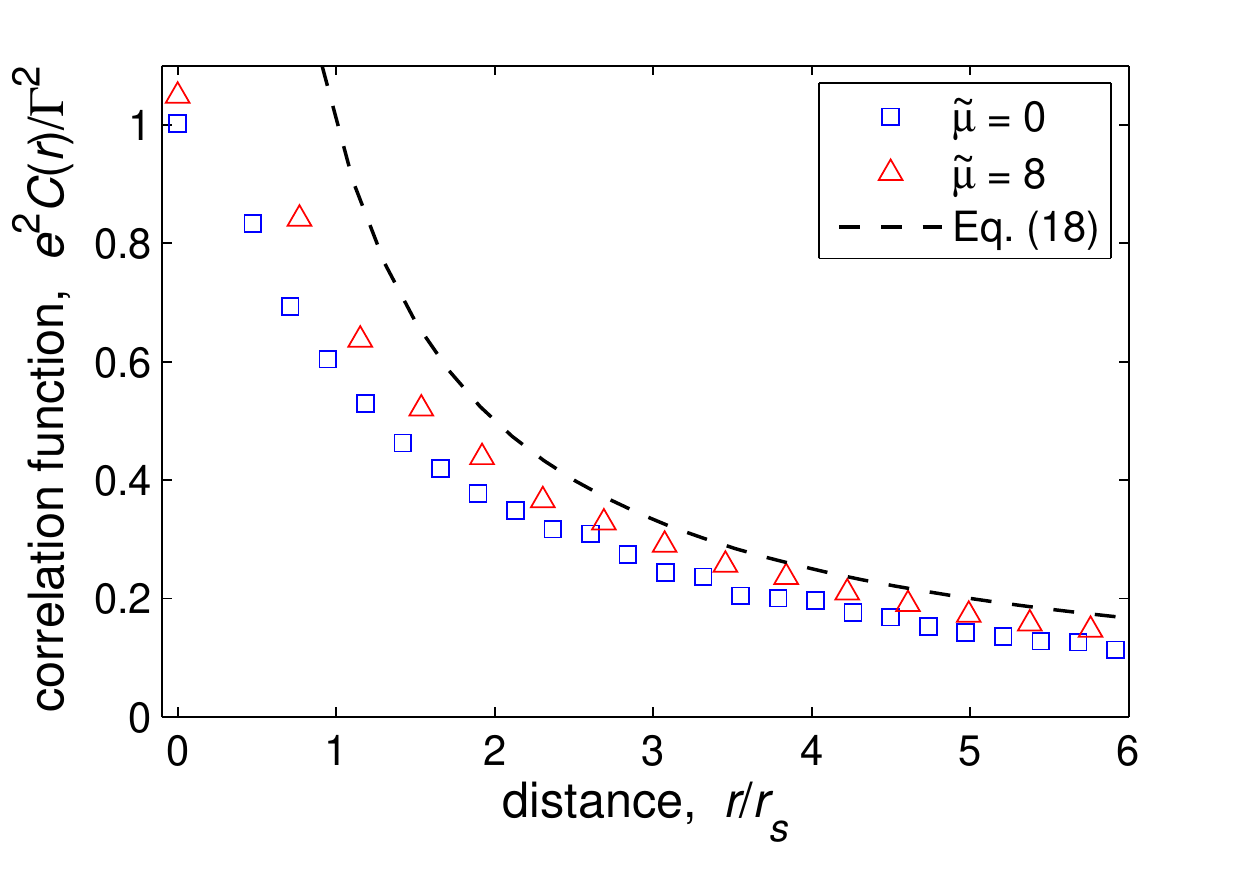}
\caption{(Color online)  The potential autocorrelation function, $C(r)$, as a function of distance.  Symbols correspond to data from our numeric simulation, both at $\mu = 0$ (squares) and at large $\mu$ (triangles), while the dashed line is our analytical theory [Eq.\ (\ref{eq:C})] for large $r/r_s$.  The vertical axis is scaled by the analytical result for $\Gamma^2$ [Eqs.\ (\ref{eq:gs0}) and (\ref{eq:gslarge})] and the horizontal axis is scaled by the analytical result for $r_s$ [Eqs.\ (\ref{eq:rs0}) and (\ref{eq:rslarge})] without fitting parameters.}
\label{fig:C}
\end{figure}

We find a solution $\phi(\rr)$ to Eq.\ (\ref{eq:selfconsistent}) by dividing the surface into a discrete grid and using numerical iteration.  Details of the iteration scheme, as well as our treatment of finite-size and finite-resolution effects, are given in the Appendix.  Once we have obtained a numerical solution to $\phi(\rr)$, the resulting variance of the disorder potential is calculated as
\be 
\Gamma^2 = e^2 \left\langle \left(\phi - \langle \phi \rangle \right)^2 \right \rangle
\label{eq:Gdef}
\ee
and the potential autocorrelation function is calculated using the definition in Eq.\ (\ref{eq:Cdef}).  All numerical results presented below are calculated at $\alpha = 0.24$ (as estimated for the experiments of Ref.\ \onlinecite{Beidenkopf2011sfh}).  Smaller $\alpha = 0.12$ was also examined, and when presented in the dimensionless units of Eqs.\ (\ref{eq:gs0}) -- (\ref{eq:rslarge}) the results were identical to those of Figs.\ \ref{fig:p2vsmu} and \ref{fig:C} to within our numerical error.

In Fig.\ \ref{fig:p2vsmu} the calculated value of $\Gt^2$ is plotted as a function of $\mt$, along with the analytical asymptotic predictions of Eqs.\ (\ref{eq:gs0}) and (\ref{eq:gslarge}).  (While Fig.\ \ref{fig:p2vsmu} presents results only for positive $\mt$, results at $\mt < 0$ are identical due to electron-hole symmetry of the Dirac point.)  The numerical results closely match the analytical theory at $\mt = 0$ and at $\mt \gg 1$.   One can notice, however, that at $\mt \approx \mt^* = 2^{2/3}$, where Eqs.\ (\ref{eq:gs0}) and (\ref{eq:gslarge}) become equal, $\Gt^2$ develops a weak maximum.  This maximum can be understood by considering that at $\mt \sim \mt^*$ the typical magnitude of the disorder potential, $\Gamma$, is similar to the typical Fermi energy, $\mu$.  As a result, screening is strongly asymmetric: positive fluctuations in potential, which increase the density of electrons, are screened more rapidly than negative fluctuations in potential, which deplete the electron density and bring the system close to the Dirac point.  The resulting distribution of the potential is skewed toward negative values of $\phi$, and this skewness produces a larger variance $\Gamma^2$ and a nonzero mean $\langle \phi \rangle$.  As $\mt$ is increased, of course, the width of the disorder potential becomes small relative to the typical Fermi energy, and screening becomes symmetric again.

In Fig.\ \ref{fig:C} we plot the potential autocorrelation function, $C(r)$, as calculated from our simulation both at zero chemical potential, $\mt = 0$, and at large chemical potential, $\mt = 8$.  In both cases the result compares well with Eq.\ (\ref{eq:C}) at large $r/r_s$ without the use of adjustable parameters.

\section{Conductivity}
\label{sec:conductivity}

In the previous sections we presented results for the disorder potential at the TI surface.  In this section we discuss the implications of our 3D model for the conductivity $\sigma$ of the surface.

In the limit of large $\mu$, where the electron density is only weakly modulated by the disorder potential, one can show using the Boltmzann kinetic equation that for electrons with a massless Dirac spectrum the conductivity is given by \cite{Culcer2010tds, Culcer2008wms, DasSarma2011eti}
\be 
\sigma = \frac{e^2}{h} \frac{\mu \tau}{4 \hbar}.
\label{eq:sigmadef}
\ee 
Here $e^2/h$ is the conductance quantum and $\tau$ is the momentum relaxation time.
In the limit of zero temperature, the scattering rate $1/\tau$ can be found by integrating the squared scattering potential produced by a given impurity over all impurities and over all scattering angles.  More specifically, one can arrive at an expression for $1/\tau$ by taking the result for the scattering rate of a 2D layer of impurities with concentration $n_i$ at distance $z$ [for example, Eq.\ (38) of Ref.\ \onlinecite{Culcer2010tds}], replacing $n_i$ with $2 N dz$, and then integrating over all planes $z$  containing impurities.  This procedure gives
\be 
\frac{1}{\tau} = \frac{k_f \alpha \kappa}{4 \pi \hbar e^2} \int_0^\infty 2 N dz \int_0^\pi d \theta \left[ \phit (2 k_f \sin\frac{\theta}{2}; z ) \right]^2 (1 - \cos^2 \theta).
\label{eq:taudef}
\ee
In this equation, $k_f = \alpha \kappa \mu / e^2$ is the Fermi wavelength, $\phit (q; z) = (2 \pi e^2/\kappa q) \exp[-q z]/[1 + (qr_s)^{-1}]$ is the screened potential (in momentum space) created by a single impurity at position $z$, and $q = 2 k_f \sin (\theta/2)$ is the change in momentum associated with scattering by an angle $\theta$.  

Evaluating the integral of Eq.\ (\ref{eq:taudef}) at small $\alpha$ gives
\be 
\frac{1}{\tau} \simeq \pi \alpha \ln \left(1/\alpha \right) \frac{e^2 N}{\hbar \kappa k_f^2}.
\ee
Inserting this result for $\tau$ into Eq.\ (\ref{eq:sigmadef}) and substituting $k_f = \sqrt{4 \pi n}$ yields the result for conductivity announced in the Introduction, Eq.\ (\ref{eq:sigman}).  This expression can also be written in terms of the (dimensionless) chemical potential as
\be 
\sigma \simeq \frac{e^2}{h} \frac{\mt^3}{4 \pi \alpha \ln(1/\alpha)},
\label{eq:sigmamu}
\ee
where $\mt = \mu/(e^2 N^{1/3}/\alpha^{2/3}\kappa)$, as defined in Sec.\ \ref{sec:theory}.

Equation (\ref{eq:sigman}) can be contrasted with the widely-used result for the 2D model of charge impurities \cite{Adam2007sct, DasSarma2011eti, Li2012tde, Culcer2010tds}, for which the conductivity is linearly proportional to the electron density: $\sigma/(e^2/h) \sim (1/\alpha^2)(n/n_i)$.  This difference can be understood conceptually by noting that, for large angle scattering, only those impurities at a distance smaller than the Fermi wavelength, $\lambda_f \sim n^{-1/2}$, contribute significantly to scattering.  One can therefore define, roughly speaking, an effective 2D concentration of scattering impurities as $N \lambda_f \sim N / n^{1/2}$.  Inserting $N/n^{1/2}$ for $n_i$ gives $\sigma \propto (1/\alpha^2)(n^{3/2}/N)$, similar to Eq.\ (\ref{eq:sigman}).  The remaining factor $1/\ln(1/\alpha)$ in Eq.\ (\ref{eq:sigman}) is related to low-angle scattering by distant impurities with $z \gg \lambda_f$.
 So far we are unaware of any transport data for TIs that shows $\sigma \propto n^{3/2}$.  Recent conductivity measurements on ultra-thin TIs (with thickness $\sim 10$\,nm $\ll \lambda_f$) suggest \cite{Kim2012sct} $\sigma \propto n$, consistent with the 2D model of impurities.

Our 3D model also produces a distinct result for the minimum conductivity $\smin$ that appears in the limit of small average electron concentration.  At small enough chemical potential that $\mt \ll 1$, the surface breaks into electron and hole puddles, and one can think that the effective carrier concentration saturates at $\sim n_p$ [see Eq.\ (\ref{eq:ne})].  An estimate of $\smin$ can therefore be obtained by setting $\mt = \mt^* = 2^{2/3}$ in Eq.\ (\ref{eq:sigmamu}), which gives the result of Eq.\ (\ref{eq:smin}).  2D models of disorder impurities also produce a minimum conductivity that is independent of the impurity concentration, but which has a different dependence on $\alpha$.  Specifically, at small $\alpha$ such models give \cite{Fogler2009npg, Adam2007sct} $\smin \sim (e^2/h) \ln(1/\alpha)$.  Our model suggests a minimum conductivity that is larger by a factor $\sim [\alpha \ln^2(1/\alpha) ]^{-1}$.


\section{Discussion}
\label{sec:discussion}

In Secs.\ \ref{sec:theory} and \ref{sec:sim} we derived analytical expressions for the magnitude of the disorder potential, the screening radius, and the autocorrelation function, and we showed that these were consistent with numerical simulations.  We now discuss the magnitude of $\Gamma$ and $r_s$ implied by these expressions for typical TIs, which generally have an impurity concentration $N \sim 10^{19}$ cm$^{-3}$.  Typical values of the Dirac velocity and fine structure constant for TIs can be taken from Ref.\ \onlinecite{Beidenkopf2011sfh}, which reports $\hbar v =  1.3$ eV\,\AA\, and estimates $\alpha = 0.24$, which corresponds to $\kappa \approx 50$ (in agreement with infrared measurements on Bi$_2$Se$_3$, for example, which yield \cite{Butch2010sss} $\kb \approx 100$).  Using these parameters gives for our unit of energy $E_0 = e^2 N^{1/3}/\kappa \alpha^{2/3} \sim 20$\,meV, and $r_0 = N^{-1/3}/\alpha^{4/3} \sim 30$\,nm.  Thus, Eqs.\ (\ref{eq:gs0}) and (\ref{eq:rs0}) imply $\Gamma \sim 30$\,meV and $r_s \sim 20$\,nm at the Dirac point, $\mu = 0$.  At large $|\mu| \gtrsim 30$\,meV, both $\Gamma^2$ and $r_s$ decay as $1/|\mu|$.

Throughout this paper we have worked within the assumption that bulk impurities are completely ionized, or in other words that there is no screening by conduction band electrons or valence band holes in the bulk.  Such an assumption is valid when the chemical potential resides in the middle of a large bulk band gap.  In this case donors or acceptors can only be neutralized by very large band bending \cite{Efros1984epo}, which brings the bulk conduction or valence band edges to the Fermi level (see, for example, Fig.\ 1 of Ref.\ \onlinecite{Skinner2012wib}).  Such fluctuations take place over a long length scale $R_g$ that scales as the square of the distance between the Fermi level and the nearest band edge.  For example, if the Fermi level is in the center of the band gap, then $R_g = E_g^2 \kappa^2/8\pi N e^4$, which is on the order of hundreds of nanometers for typical TIs \cite{Skinner2012wib}.  On the other hand, near the surface of the TI the potential fluctuations are screened much more effectively and over a much shorter distance, $r_s$, by the (ungapped) surface states.  As shown above, $r_s$ is typically $\lesssim 20$\,nm, and the amplitude of surface potential fluctuations $\Gamma \sim 30$\,meV $\ll E_g \sim 300$\,meV.  One can therefore safely assume that near the surface there is no large band bending and one can indeed treat bulk impurities as completely ionized.  The effect of bulk screening should appear only in the long-range behavior of the correlation function, $r \gg R_g$, where the $1/r$ decay of $C(r)$ is truncated and, as one can show, is replaced with $C(r) \sim e^2 N R_g r_s^2/\kappa^2 r^2$.

We now compare our results for $\Gamma$ and $r_s$ to the recent experiments of Ref.\ \onlinecite{Beidenkopf2011sfh}.  Those authors examined samples of doped Bi$_2$Te$_3$ and Bi$_2$Se$_3$ for which $\mu \sim 100$\,meV, and they found that the electric potential at the surface was well-characterized by a Gaussian distribution with a width $\sim 20$ -- $40$\,meV, which corresponds to $\Gamma \sim 10$ -- $20$\,meV.  The characteristic length scale of potential fluctuations was measured as $r_s \sim 20$ -- $30$\,nm.  Using the estimate above for $E_0$ suggests that for these samples $\mt \sim 6$.  Equations (\ref{eq:gslarge}) and (\ref{eq:rslarge}) then give $\Gamma \sim 18$\,meV and $r_s \sim 5$\,nm, which is in reasonable agreement with experiment.  Further, evidence from Ref.\ \onlinecite{Beidenkopf2011sfh} suggests that the surface disorder potential originates primarily from impurities deep below the TI surface, and that its magnitude is relatively independent of the type of (monovalent) impurity present.  These findings are again consistent with the theory presented here.
So far, measurements of the potential autocorrelation function $C(r)$ have not been reported, but they can in principle be extracted from the measurements of Ref.\ \onlinecite{Beidenkopf2011sfh} and compared to our prediction above.  

It is also worth mentioning that the theory we have presented here can be applied to graphene on a substrate having large dielectric constant (so that $\alpha = e^2/\kappa \hbar v \ll 1$) and embedded bulk impurity charges.  For this application one should only replace $\nu$ everywhere with $4 \nu$, since graphene's spin and valley degeneracy give it a four times larger density of states at a given energy.  This substitution produces values of $\Gamma^2$ and $r_s$ for graphene that are four times smaller than what is written in Secs.\ \ref{sec:intro} and \ref{sec:theory}.

Finally, we note that our theory ignores the possibility of screening by material \emph{outside} the TI.  For example, if the TI is placed next to a metal electrode or an ionic liquid \cite{Xiong2012tqo}, then this external material can screen the large potential fluctuations created by the bulk, thereby decreasing $\Gamma$ and $r_s$.

\acknowledgments

The authors are grateful to Y. Ando, H. Beidenkopf, M. M. Fogler, Q. Li, and A. Yazdani for helpful discussions.  This work was supported primarily by the National Science Foundation through the University of Minnesota MRSEC under Award Number DMR-0819885.

\appendix

\section{Numerical solution of the simulated self-consistent potential}

In Sec.\ \ref{sec:sim} we presented results from a numeric solution of the potential at a simulated TI surface.  In this Appendix we discuss the details of our numerical method, including our iteration scheme and our treatment of finite size and finite resolution effects.

In our simulation, $NL^3$ impurities, each with a random sign and random position, are placed in a volume with dimensions $L \times L \times L/2$ and open boundary conditions.  One of the square faces of this volume is divided into a square grid with $\rho (L+1)^2$ grid points, where $\rho$ is the grid resolution.  In order to solve numerically for the potential at this surface for a given value of the parameters $\mu$, $L$, and $\rho$, we use a numerical iteration scheme that makes successive approximations $\phi^{(n)}(\rr)$ for the potential at each grid point $\rr$ using Eq.\ (\ref{eq:selfconsistent}).  The first approximation is made by evaluating the bare potential $\pext(\rr)$ at each grid point $\rr$, which we equate with $\phi^{(0)}(\rr)$.  We then evaluate the right hand side of Eq.\ (\ref{eq:selfconsistent}) for each $\rr$, which we denote $\pnew^{(0)}(\rr)$, by inserting $\phi^{(0)}(\rr)$ for $\phi(\rr)$.  Rather than setting $\phi^{(1)}(\rr) = \pnew^{(0)}(\rr)$ directly, we use a standard underrelaxation scheme with a damping parameter $\gamma$ to improve convergence of the solution: 
\be 
\phi^{(n+1)}(\rr) = \gamma \phi^{(n)}(\rr) + (1 - \gamma)\pnew^{(n)}(\rr).
\label{eq:it}
\ee
This process is continued iteratively, with $\pnew^{(n)}(\rr)$ evaluated at each iteration by inserting $\phi^{(n)}(\rr)$ into Eq.\ (\ref{eq:selfconsistent}) and used to create a revised estimate $\phi^{(n+1)}(\rr)$ according to Eq.\ (\ref{eq:it}).  The iteration is halted when the value of $\Gamma^2$ associated with $\phi^{(n)}(\rr)$ [see Eq.\ (\ref{eq:Gdef})] has converged to within $0.01\%$.  For each value of the simulation parameters, all results are averaged over $100$ random placements of the bulk impurity charges.  Results presented above use $\gamma = 0.5$, but we verified that our convergent solution is independent of the value of $\gamma$ chosen for $0.2 < \gamma < 0.98$.

Because of the long-ranged nature of the potential fluctuations created by bulk impurities, finite size effects in these simulations can be significant.  The ideal numerical result corresponds to the limit of an infinitely large simulation volume with an infinitely well-resolved spatial grid, $L \rightarrow \infty$ and $\rho \rightarrow \infty$. In practice, approaching this limit very closely can require an unrealistically large simulation.  We therefore make use of an extrapolation method to estimate the value of $\Gamma^2$ corresponding to the $L \rightarrow \infty$ and $\rho \rightarrow \infty$ limit.  Specifically, we find that for a given value of the chemical potential $\mu$ and grid resolution $\rho$, the variance in the potential can be well fitted to the equation $\Gamma^2(L) = \Gamma^2_\infty - A/L$, where $A$ is some positive constant.   This dependence can be justified theoretically by considering that the perimeter of the simulated surface, which contains a fraction $\propto 1/L$ of the total number of grid points, has a smaller amplitude of the potential than the center of the grid by virtue of being at the edge of the simulation volume.  The value of $\Gamma^2_\infty$ is then extracted by making a best fit to $\Gamma^2(L)$ using a range of simulated sizes.  Typically, our simulations use $L N^{1/3} = 20, 25, 30, 40, 50, 60$  and we see a coefficient of determination $R^2 > 0.95$ for the linear fit of $\Gamma^2$ as a function of $1/L$.  (For $\alpha = 0.24$ the theoretical screening radius at $\mu = 0$ is $r_s \approx 4.2 N^{-1/3}$.)

A similar extrapolation is also performed to evaluate the limit $\rho \rightarrow \infty$.  The $(1/L)$-extrapolated values of $\Gamma^2$ at a given grid density $\rho$ are fitted to a linear function of $1/\rho$, and the final estimate of $\Gamma^2$ for a given $\mu$ is equated with the $y$-intercept of the corresponding line.  The result of both of these extrapolations produces an estimate for $\Gamma^2$ that is at most $11\%$ different from the ones taken directly from our largest simulated sizes.  Larger $\Gamma^2$ generally corresponds to larger extrapolation.

The correlation function $C(r)$ plotted in Fig.\ \ref{fig:C} is not the result of an extrapolation, but is calculated directly from a single set of simulations with large $L \approx 50 r_s$ and $\rho \approx (4/r_s)^2$, averaged over $100$ random placements of the bulk impurity charges for each value of $\mt$.

\bibliography{surface_potential}

\end{document}